\documentclass{PoS}
\usepackage{subfigure}
\title{Towards a unitary Dalitz plot analysis of three-body hadronic $B$ decays
\thanks{This work has been supported in part by the Polish Ministry of Science
and Higher Education (grant No N N202 248135) and by the IN2P3-Polish
Laboratories Convention (project No 08-127).}}

\ShortTitle{Towards a unitary Dalitz plot analysis}

\author{\speaker{L.~Le\'sniak$^{a}$}\\
$^{ a}$ Division of Theoretical Physics, The Henryk Niewodnicza\'nski
Institute of Nuclear Physics, Polish Academy of Sciences, 31-342 Krak\'ow,
Poland
\\
        E-mail: \email{Leonard.Lesniak@ifj.edu.pl}}

\author{B.~El-Bennich$^{ b}$, A.~Furman$^{ c}$, R.~Kami\'nski$^{ a}$,
 B.~Loiseau$^{ d}$,
B.~Moussallam$^{ e}$\\
$^{ b}$ Physics Division, Argonne National Laboratory, Argonne, Illinois,
 60439, USA\\
$^{ c}$ ul. Bronowicka 85/26, 30-091 Krak\'ow, Poland \\
$^{ d}$ Laboratoire de Physique Nucl\'eaire et de Hautes \'Energies 
(IN2P3--CNRS--Universit\'es Paris 6 ~~~~et 7), Groupe Th\'eorie,
Universit\'e Pierre et Marie Curie, 4 place Jussieu, 75252 Paris, France
\thanks{Unit\'e de Recherche des Universit\'es
Paris 6 et Paris 7, associ\'ee au CNRS} \\
$^{ e}$ Groupe de Physique Th\'eorique, Institut de Physique Nucl\'eaire 
(IN2P3--CNRS),
Universit\'e Paris-Sud 11, 91406 Orsay Cedex, France
}

\abstract{A unitary model of the final state $K \pi$ interaction amplitudes in the 
$B \to K \pi\pi$ decays is constructed. The weak decay penguin amplitudes,
derived in QCD factorization, are supplemented by phenomenological contributions.
The strange $K \pi$ scalar and vector form factors are used to calculate the $K \pi$
effective mass and helicity angle distributions, branching ratios, 
CP asymmetries and the phase difference between the $B^0$ and $\bar B^0$ decay 
amplitudes to $K^*(892) \pi$. The fit on the phenomenological parameters   
leads to a good agreement with the experimental data, particularly for the 
$B \to K^*(892) \pi$ decays. However, our predicted 
$B^\pm \to \ K^*_0(1430)\pi^\pm$,  $K^*_0(1430) \to K^\pm\pi^\mp$ branching 
fraction is smaller than the results
of the Belle and BaBar collaborations, obtained from isobar model analyses.
A new parameterization of the $S$-wave $K \pi$ effective mass distribution, which
can be used in future experimental Dalitz plot analyses, is proposed.
}

\FullConference{The 2009 Europhysics Conference on High Energy Physics,\\
		 July 16 - 22 2009\\
		 Krakow, Poland}

\begin{document}

\section{Introduction}Studies of three-body charmless hadronic decays of $B$
mesons are very useful not only in standard model tests and in searches for 
"new
 physics" effects but also in the determination of strong interaction amplitudes.
 Both weak and strong interactions can create structures seen
 on Dalitz plots. Analyses of these diagrams should
be performed in a unitary approach which allows for
a proper construction of $B$-decay amplitudes. Usually a suitable partial wave 
analysis of final state amplitudes should be done. Then an adequate 
determination of branching fractions and CP asymmetries for different quasi-two 
body decay reactions is possible.

Construction of a fully unitary three-body strong interaction amplitude is, 
however, a difficult task. A first step towards this goal is to
enforce two-body unitarity. Here we apply this concept to the $K \pi$ channel 
for the $B \to K \pi\pi$  decays by using a unitary coupled channel model.
We study the $K \pi$ amplitudes in the 
limited $K \pi$ effective mass range smaller than about 1.8 GeV. Our aim is to
describe such physical quantities as differential effective mass and helicity
 angle distributions, integrated branching fractions and 
direct CP asymmetries.

In experimental analyses of $B$-decays the isobar model
is very frequently applied. Within that model quasi-two body branching fractions
 are determined.
However, the decay amplitudes  commonly used in the  isobar 
model are not  unitary neither in three-body decay channels nor
in two-body subchannels. This lack of unitarity can create severe problems in 
the determination of 
branching fractions in the case of wide overlapping resonances. In $B \to K \pi\pi$
decays one observes a wide $S$-wave resonance $K^*_0(1430)$. Its width equals to
about 270 MeV and the postulated  $K^*_0(800)$ state can have even larger 
width of 500 MeV. Thus an important source of model errors in extraction of  
the branching  ratio for the decay $B^\pm \to \ K^*_0(1430)\pi^\pm$
is a possible wrong attribution of a part of Dalitz plot density to a background
amplitude and to  its interference with other amplitudes, mostly with the
$S$-wave. Below, we shall briefly discuss that issue.
\section{Theoretical model}
The weak decay amplitudes of $B^{\pm}$, $B^0$ and $\bar B^0$,  which are 
derived in QCD 
factorization, are supplemented by phenomenological contributions to the penguin
amplitudes. Strong interaction
amplitudes are constrained by chiral symmetry, QCD and experimental data on
meson-meson interactions. The matrix elements of the effective weak Hamiltonian
involve the strange $K \pi$ scalar and vector form factors. The introduction of
form factors, constrained by  theory and other experiments than $B$ decays, is 
an alternative to the use of the isobar model. The $K \pi$ $S$-wave contribution
to the $B^- \to K^- \pi^+\pi^-$ amplitude reads:
 
\begin{eqnarray*}
\label{K-pi+Sampli}
\mathcal{M}_S^-\equiv \langle \pi^-\ (K^-\pi^+)_S\vert H_{eff}\vert B^-\rangle
=\frac{G_F}{\sqrt{2}} (M_B^2-m_{\pi}^2)\frac{m_{K}^2-m_{\pi}^2}{q^2}
f_0^{B^-\pi^-}(q^2) \ f_0^{K^-\pi^+}(q^2)\\
\times
\bigg\{
\lambda_u\left(a_4^{u}(S)-\frac{a_{10}^{u}(S)}{2}+c_4^{u}\right)
+\lambda_c\left(a_4^c(S)-\frac{a_{10}^c(S)}{2}+c_4^{c}\right)\\
 -\ \frac{2q^2}{(m_b-m_d)(m_s-m_d)} \left[
\lambda_u\left(a_6^{u}(S)-\frac{a_8^{u}(S)}{2}+c_6^{u}\right)
+\lambda_c\left(a_6^c(S)-\frac{a_8^c(S)}{2}+c_6^{c}\right)
\right]\bigg\}.
\end{eqnarray*}
Expressions for other contributions to the $B$ decay amplitudes can be found in
~\cite{El-Bennich2006}. In the above equation
 $G_F$ denotes the Fermi coupling constant, $M_B,\ m_K$ and $m_\pi$ are the
 masses of the charged $B$ mesons, 
kaons and pions, $f_0^{B^-\pi^-}(q^2)$ and $f_0^{K^-\pi^+}(q^2)$
are the $B^-$ to $\pi^-$ and the $K\pi$ scalar form factors. The symbols 
$\lambda_u= V_{ub} V^*_{us}, \lambda_c= V_{cb} V^*_{cs}$, are products of the 
Cabibbo-Kobayashi-Maskawa quark-mixing matrix elements $V_{qq'}$; $a^{u,c}_j(S)$,
$j=4,6,8,10$ are the coefficients of the effective Hamiltonian $H_{eff}$, $q^2$ 
is the $K\pi$ effective mass squared and $m_b,\ m_s$ and $\ m_d$ are $b,~ s$ and
 $d$ quark
masses. Finally the  $c_4^{u},c_4^{c}, c_6^{u}$ and  $c_6^{c}$ are the
phenomenological complex parameters which are fitted to the experimental data.

The scalar and vector $K\pi$ form factors are connected to scattering amplitudes 
in the $S$ and $P$ waves via unitarity relations. In the $S$ wave 
we treat two coupled $ K\pi$ and $K\eta '$ amplitudes. 
Three coupled channels: $ K\pi$, $K^*\pi$ and $K\rho$ appear in the
$P$ wave. The corresponding scattering amplitudes are constrained by experimental data,
especially by the LASS results obtained at SLAC. The Muskhelishvili-Omnes
equations are used to calculate the $K \pi$ form factors. 
\section{Results and discussion}
 The $ K\pi$ effective mass and helicity angle distributions, 
branching ratios, CP asymmetries and the phase difference between the $B^0$ and
$\bar B^0$ decay amplitudes to $K^*(892)\pi$ are calculated using the
minimization program which serves to fit the set of 319 data from the Belle and
BaBar collaborations with the four phenomenological parameters $c_{4,6}^{u,c}$.
As input 
we use only the well measured branching fractions for the ${B} \to K^*(892)\pi$.
We do not use the experimental branching fractions for the 
${B} \to K^*_0(1430)\pi$, which
 are not well determined due to the large width of the $K^*_0(1430)$ resonance.
Some of our model predictions for the $S$-wave part of the branching fraction
for the $B \to K \pi\pi$ decays are given in Table 1. One can notice that they
are lower than the experimental data, being substantially smaller than the 
Belle results and closer to the BaBar numbers. 
If the parameters $c_{4,6}^{u,c}$ are all put equal to zero then the  $P$-wave part of 
the branching fraction is underestimated by a factor of 4 to 5 and 
the $S$-wave part by a factor of 2. 
 The fit on the model parameters leads to a good agreement with the experimental data, particularly 
for the kaon-pion effective mass and helicity angle distributions. Some results
are presented in Fig.1. 

Based on a good description of data we propose to choose in the future
experimental analyses the following parameterization of the $S$-wave 
$B \to K \pi\pi$ amplitude:
\begin{equation}
\mathcal{M}_S^- =  f_0^{K\pi}(m_{K\pi}^2)(c_0/m_{K\pi}^2 + c_1),
\nonumber
\end{equation}
where $f_0^{K\pi}$ is the scalar $K\pi$ form factor while $c_0$ and $c_1$ are
complex numbers to be fitted from the data. Numerical values of the complex scalar form factor can be provided
 on request.
 \newpage
 \begin{table*}[t]
\begin{center}
\caption{Branching ratios averaged over charge conjugate reactions 
$B\to K\pi\pi$ in units of $10^{-6}$. $(K^+\pi^-)_P$ and $(K^+\pi^-)_S$ denote the
$(K^+\pi^-)$ pair in $P$-wave and $S$-wave, respectively. In the first two lines, 
the values of the model, calculated by the integration over the given 
$m_{K\pi}$ range,
 are compared to the corresponding Belle ~\cite{Garmash:2005rv,Garmash2007} and 
BaBar ~\cite{Aubert:2008bj,Aubert:2007bs} results written in the third and fourth 
column. In the last two lines,
the Belle branching fractions~\cite{Garmash:2005rv,Garmash2007}, calculated with
a $K^*_0(1430)$ Breit-Wigner amplitude, and the BaBar branching 
fractions~\cite{Aubert:2008bj,Aubert:2007bs}, calculated in  their 
parametrization of the $K\pi$ $S$-wave, are compared to our model predictions. 
 The model errors are the phenomenological parameter uncertainties 
 found in the minimization procedure.
}
\medskip
\begin{tabular}[]{|c|c|c|c|c|}
\hline
	decay mode   & $m_{K\pi}$ range (GeV) & Belle          &  BaBar       &  model\\

\hline
 $B^+\to (K^+\pi^-)_P\ \pi^-$ &$(0.82,0.97)$   & $5.35\pm0.59$  & $5.98\pm0.75$ &
 $5.73\pm0.14$ \\

\hline
 $B^0\to (K^0\pi^+)_P\ \pi^-$ &$(0.82,0.97)$   & $4.65\pm0.77$  & $6.47\pm0.75$ &
 $5.42\pm0.16$ \\
 
\hline
 $B^+\to (K^+\pi^-)_S\ \pi^-$ &$(0.64,1.76)$   & $27.0\pm2.5$  & $22.5\pm4.6$ &
 $16.5\pm0.8$ \\

\hline
 $B^0\to (K^0\pi^+)_S\ \pi^-$ &$(0.64,1.76)$  & $26.0\pm3.4$  & $17.3\pm4.6$ &
 $15.8\pm0.7$ \\

\hline

\end{tabular}
\end{center}
\end{table*}
\medskip
\begin{figure}[!ht]
\subfigure{\includegraphics[width=0.45\textwidth]{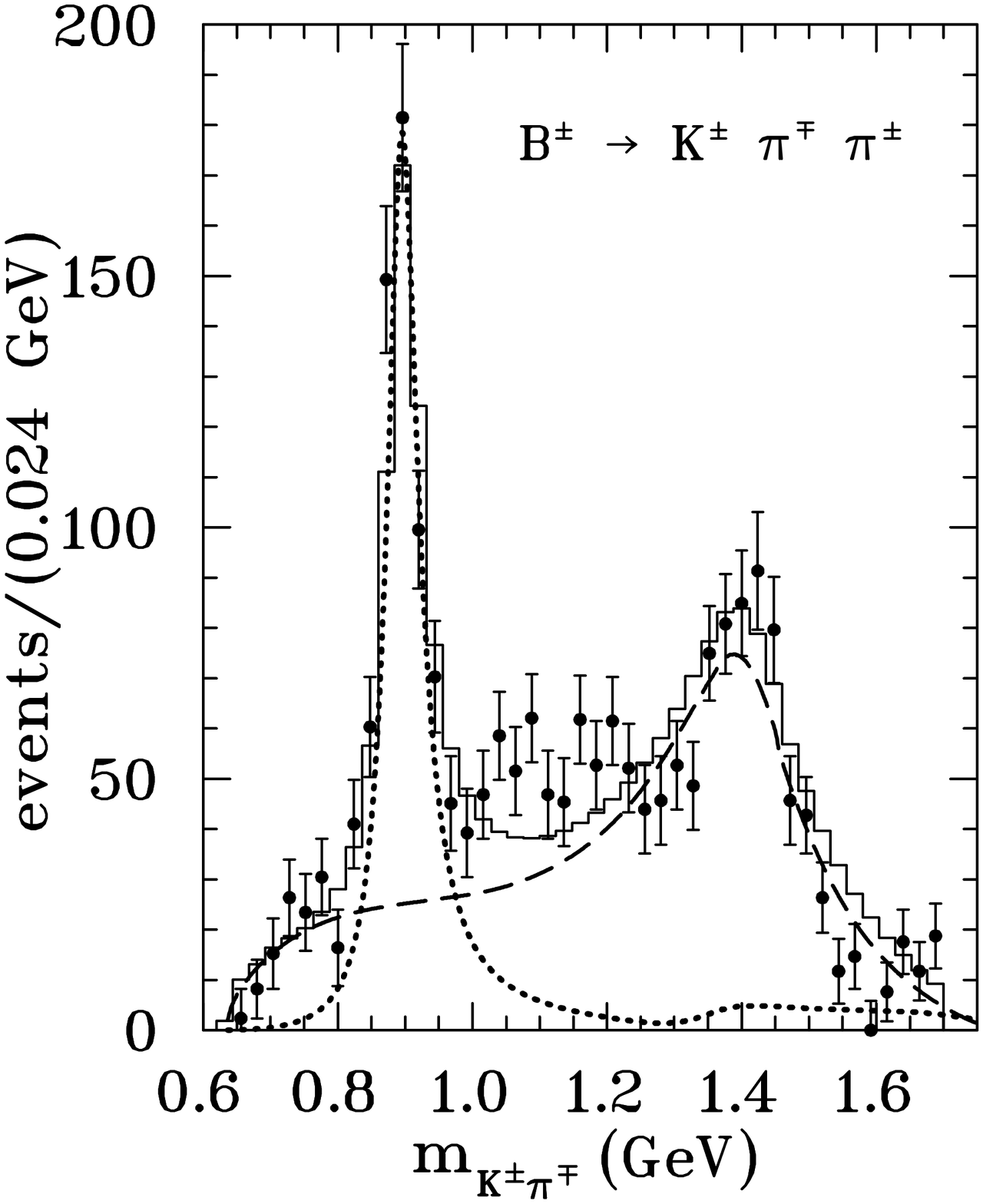}}~~~~~
\subfigure{\includegraphics[width=0.465\textwidth]{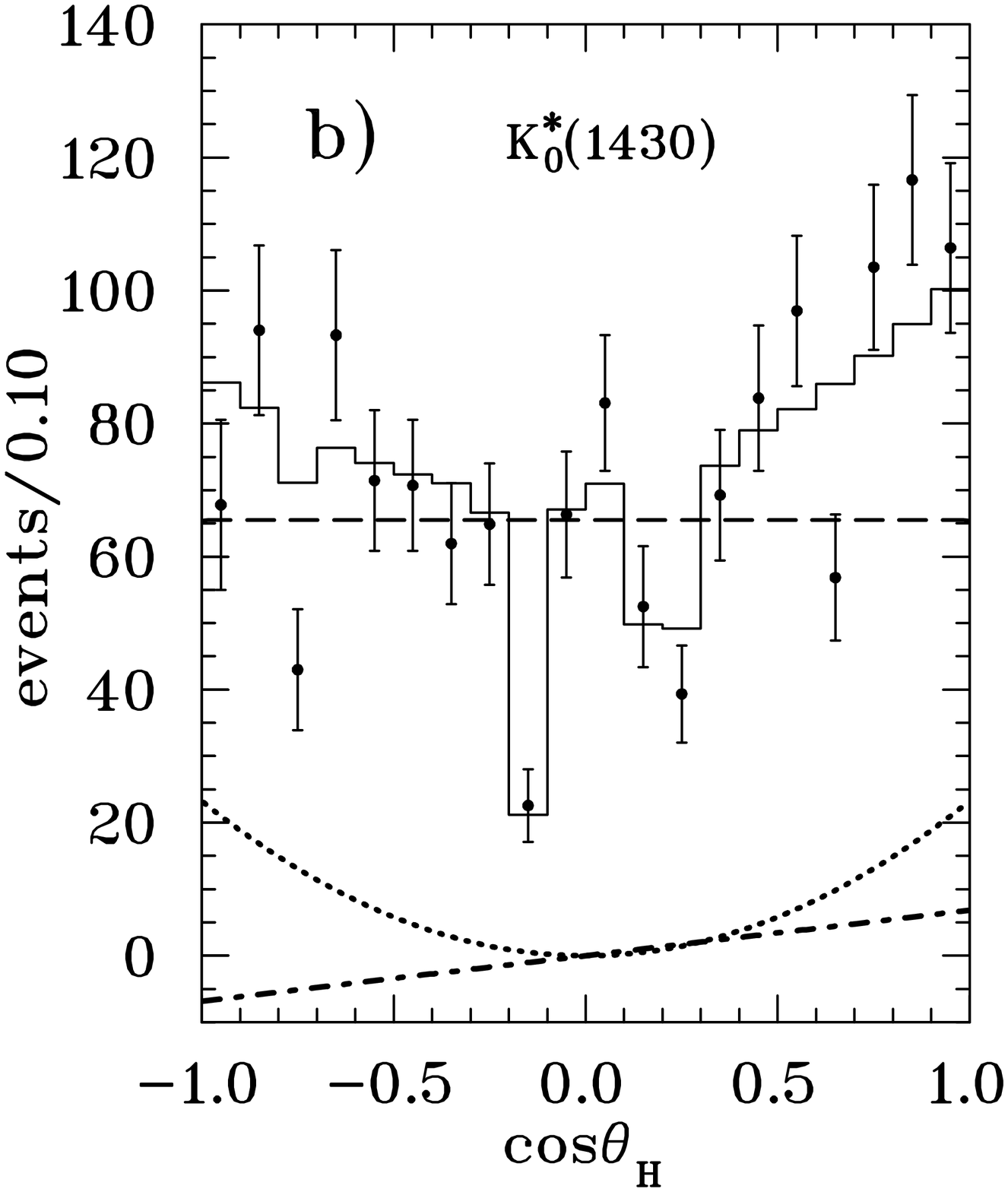}}
\caption{a) The  $K^\pm \pi^\mp$ effective mass distributions in the 
 $B^\pm \to K^\pm \pi^\mp \pi^\pm$ decays.
 Data points are from   Ref.~\cite{Aubert:2008bj}. The
dashed line represents the $S$-wave contribution of our model, the dotted line 
that of the $P$-wave and  the  histogram corresponds to the coherent sum of the
 $S$- and $P$-wave contributions.
b) Helicity angle distribution for  $B^\pm \to K^\pm \pi^\mp \pi^\pm$ decays 
 calculated from the averaged double differential distribution  integrated over
  $m_{K^\pm \pi^\mp}$ mass from 1.0 to 1.76 GeV.  
 Data  points are from  Ref.~\cite{Abe2005}. 
The dashed line represents the $S$-wave contribution of our model, the dotted 
line that
 of the $P$-wave and the dot-dashed line that of the interference term.
The histogram corresponds to the sum of these three contributions.
 }
\end{figure}


\vspace{-12.0cm}

\hspace{1cm} {\Large{ a)}}


\end{document}